%Paper: cond-mat/9512175
%From: "Prof. Hiroshi KURATSUJI" <kra@bkc.ritsumei.ac.jp>
%Date: Fri, 29 Dec 1995 17:55:25 +0900
%Date (revised): Fri, 29 Dec 1995 18:50:44 +0900

%%%%%%%%%%%%%%%%%%%%%%%%%%%%%%%%%%%%%%%%%%%%%%%%%%%%%%%%%%%%%%%%%%%%%%%%
% % % % % % % % % % % % % % % % % % % % % % % % % % % % % % % % % % % %
%%%   This is PHYZZX macro package.   % % % % % % % % % % % % % % % % %
%% % % % % % % % % % % % % % % % % % % % % % % % % % % % % % % % % % % %
%%%  This version of PHYZZX should be used with Version 1.0 of TEX  % %
%% % % % % % % % % % % % % % % % % % % % % % % % % % % % % % % % % % % %
%%%   Do not "\input phyzzx" unless you preload or "\input" PLAIN.  % %
%% % % % % % % % % % % % % % % % % % % % % % % % % % % % % % % % % % % %
%%%   To preload both PLAIN and PHYZZX, begin your file with	% % % %
%%%  a line "%macropackage=phyzzx" instead of "\input phyzzx".  % % % %
%% % % % % % % % % % % % % % % % % % % % % % % % % % % % % % % % % % % %
%%%%%%%%%%%%%%%%%%%%%%%%%%%%%%%%%%%%%%%%%%%%%%%%%%%%%%%%%%%%%%%%%%%%%%%%
%%%%%%%  Created by Vadim Kaplunovsky in June 1984.   %%%%%%%%%%%%%%%%%%
% % % % % % % % % % % % % % % % % % % % % % % % % % % % % % % % % % % %
%%%%%%%%%%%%  Latest update/debug: September 24, 1984.	  %%%%%%%%%%%%%%
%%%%%%%%%%%%%%%%%%%%%%%%%%%%%%%%%%%%%%%%%%%%%%%%%%%%%%%%%%%%%%%%%%%%%%%%
%
\catcode`@=11 % This allows us to modify PLAIN macros.
%
%%%%%%%%%%%%%%%%%%%%%%%%%%%%%%%%%%%%%%%%%%%%%%%%%%%%%%%%%%%%%%%%%%%%%%%%
%
%   I begin with fonts.
%

\font\fourteenrm=cmr10 scaled\magstep2
\font\twelverm=cmr10 scaled\magstep1
\font\ninerm=cmr9	     \font\sixrm=cmr6

\font\fourteenbf=cmbx10 scaled\magstep2
\font\twelvebf=cmbx10 scaled\magstep1
\font\ninebf=cmbx9	      \font\sixbf=cmbx6
\font\seventeeni=cmmi10 scaled\magstep3	    \skewchar\seventeeni='177
\font\fourteeni=cmmi10 scaled\magstep2	    \skewchar\fourteeni='177
\font\twelvei=cmmi10 scaled\magstep1	    \skewchar\twelvei='177
\font\ninei=cmmi9			    \skewchar\ninei='177
\font\sixi=cmmi6			    \skewchar\sixi='177
\font\seventeensy=cmsy10 scaled\magstep3    \skewchar\seventeensy='60
\font\fourteensy=cmsy10 scaled\magstep2	    \skewchar\fourteensy='60
\font\twelvesy=cmsy10 scaled\magstep1	    \skewchar\twelvesy='60
\font\ninesy=cmsy9			    \skewchar\ninesy='60
\font\sixsy=cmsy6			    \skewchar\sixsy='60

\font\fourteenex=cmex10 scaled\magstep2
\font\twelveex=cmex10 scaled\magstep1

\font\fourteensl=cmsl10 scaled\magstep2
\font\twelvesl=cmsl10 scaled\magstep1
\font\ninesl=cmsl9

\font\fourteenit=cmti10 scaled\magstep2
\font\twelveit=cmti10 scaled\magstep1
\font\twelvett=cmtt10 scaled\magstep1
\font\twelvecp=cmcsc10 scaled\magstep1
\font\tencp=cmcsc10
\newfam\cpfam
%
	% quick fix for a missing font
%
\newcount\f@ntkey	     \f@ntkey=0
\def\samef@nt{\relax \ifcase\f@ntkey \rm \or\oldstyle \or\or
	 \or\it \or\sl \or\bf \or\tt \or\caps \fi }
\def\fourteenpoint{\relax
    \textfont0=\fourteenrm	    \scriptfont0=\tenrm
    \scriptscriptfont0=\sevenrm
     \def\rm{\fam0 \fourteenrm \f@ntkey=0 }\relax
    \textfont1=\fourteeni	    \scriptfont1=\teni
    \scriptscriptfont1=\seveni
     \def\oldstyle{\fam1 \fourteeni\f@ntkey=1 }\relax
    \textfont2=\fourteensy	    \scriptfont2=\tensy
    \scriptscriptfont2=\sevensy
    \textfont3=\fourteenex     \scriptfont3=\fourteenex
    \scriptscriptfont3=\fourteenex
    \def\it{\fam\itfam \fourteenit\f@ntkey=4 }\textfont\itfam=\fourteenit
    \def\sl{\fam\slfam \fourteensl\f@ntkey=5 }\textfont\slfam=\fourteensl
    \scriptfont\slfam=\tensl
    \def\bf{\fam\bffam \fourteenbf\f@ntkey=6 }\textfont\bffam=\fourteenbf
    \scriptfont\bffam=\tenbf	 \scriptscriptfont\bffam=\sevenbf
    \def\tt{\fam\ttfam \twelvett \f@ntkey=7 }\textfont\ttfam=\twelvett
    \h@big=11.9\p@{} \h@Big=16.1\p@{} \h@bigg=20.3\p@{} \h@Bigg=24.5\p@{}
    \def\caps{\fam\cpfam \twelvecp \f@ntkey=8 }\textfont\cpfam=\twelvecp
    \setbox\strutbox=\hbox{\vrule height 12pt depth 5pt width\z@}
    \samef@nt}
\def\twelvepoint{\relax
    \textfont0=\twelverm	  \scriptfont0=\ninerm
    \scriptscriptfont0=\sixrm
     \def\rm{\fam0 \twelverm \f@ntkey=0 }\relax
    \textfont1=\twelvei		  \scriptfont1=\ninei
    \scriptscriptfont1=\sixi
     \def\oldstyle{\fam1 \twelvei\f@ntkey=1 }\relax
    \textfont2=\twelvesy	  \scriptfont2=\ninesy
    \scriptscriptfont2=\sixsy
    \textfont3=\twelveex	  \scriptfont3=\twelveex
    \scriptscriptfont3=\twelveex
    \def\it{\fam\itfam \twelveit \f@ntkey=4 }\textfont\itfam=\twelveit
    \def\sl{\fam\slfam \twelvesl \f@ntkey=5 }\textfont\slfam=\twelvesl
    \scriptfont\slfam=\ninesl
    \def\bf{\fam\bffam \twelvebf \f@ntkey=6 }\textfont\bffam=\twelvebf
    \scriptfont\bffam=\ninebf	  \scriptscriptfont\bffam=\sixbf
    \def\tt{\fam\ttfam \twelvett \f@ntkey=7 }\textfont\ttfam=\twelvett
    \h@big=10.2\p@{}
    \h@Big=13.8\p@{}
    \h@bigg=17.4\p@{}
    \h@Bigg=21.0\p@{}
    \def\caps{\fam\cpfam \twelvecp \f@ntkey=8 }\textfont\cpfam=\twelvecp
    \setbox\strutbox=\hbox{\vrule height 10pt depth 4pt width\z@}
    \samef@nt}
\def\tenpoint{\relax
    \textfont0=\tenrm	       \scriptfont0=\sevenrm
    \scriptscriptfont0=\fiverm
    \def\rm{\fam0 \tenrm \f@ntkey=0 }\relax
    \textfont1=\teni	       \scriptfont1=\seveni
    \scriptscriptfont1=\fivei
    \def\oldstyle{\fam1 \teni \f@ntkey=1 }\relax
    \textfont2=\tensy	       \scriptfont2=\sevensy
    \scriptscriptfont2=\fivesy
    \textfont3=\tenex	       \scriptfont3=\tenex
    \scriptscriptfont3=\tenex
    \def\it{\fam\itfam \tenit \f@ntkey=4 }\textfont\itfam=\tenit
    \def\sl{\fam\slfam \tensl \f@ntkey=5 }\textfont\slfam=\tensl
    \def\bf{\fam\bffam \tenbf \f@ntkey=6 }\textfont\bffam=\tenbf
    \scriptfont\bffam=\sevenbf	   \scriptscriptfont\bffam=\fivebf
    \def\tt{\fam\ttfam \tentt \f@ntkey=7 }\textfont\ttfam=\tentt
    \def\caps{\fam\cpfam \tencp \f@ntkey=8 }\textfont\cpfam=\tencp
    \setbox\strutbox=\hbox{\vrule height 8.5pt depth 3.5pt width\z@}
    \samef@nt}
%
%%%%%%%%%%%%%%%%%%%%%%%%%%%%%%%%%%%%%%%%%%%%%%%%%%%%%%%%%%%%%%%%%%%%%%%%
%
%   Next redifine \big \Big \bigg and \Bigg to work with all fonts
%
%%%%%%%%%%%%%%%%%%%%%%%%%%%%%%%%%%%%%%%%%%%%%%%%%%%%%%%%%%%%%%%%%%%%%%%%
%
\newdimen\h@big  \h@big=8.5\p@
\newdimen\h@Big  \h@Big=11.5\p@
\newdimen\h@bigg  \h@bigg=14.5\p@
\newdimen\h@Bigg  \h@Bigg=17.5\p@
\def\big#1{{\hbox{$\left#1\vbox to\h@big{}\right.\n@space$}}}
\def\Big#1{{\hbox{$\left#1\vbox to\h@Big{}\right.\n@space$}}}
\def\bigg#1{{\hbox{$\left#1\vbox to\h@bigg{}\right.\n@space$}}}
\def\Bigg#1{{\hbox{$\left#1\vbox to\h@Bigg{}\right.\n@space$}}}
%
%%%%%%%%%%%%%%%%%%%%%%%%%%%%%%%%%%%%%%%%%%%%%%%%%%%%%%%%%%%%%%%%%%%%%%%%
%
%   Next, I define basic spacing parameters.
%
\normalbaselineskip = 20pt plus 0.2pt minus 0.1pt
\normallineskip = 1.5pt plus 0.1pt minus 0.1pt
\normallineskiplimit = 1.5pt
\newskip\normaldisplayskip
\normaldisplayskip = 20pt plus 5pt minus 10pt
\newskip\normaldispshortskip
\normaldispshortskip = 6pt plus 5pt
\newskip\normalparskip
\normalparskip = 6pt plus 2pt minus 1pt
\newskip\skipregister
\skipregister = 5pt plus 2pt minus 1.5pt
\newif\ifsingl@	   \newif\ifdoubl@
\newif\iftwelv@	   \twelv@true
\def\singlespace{\singl@true\doubl@false\spaces@t}
\def\doublespace{\singl@false\doubl@true\spaces@t}
\def\normalspace{\singl@false\doubl@false\spaces@t}
\def\Tenpoint{\tenpoint\twelv@false\spaces@t}
\def\Twelvepoint{\twelvepoint\twelv@true\spaces@t}
\def\spaces@t{\relax%
 \iftwelv@ \ifsingl@\subspaces@t3:4;\else\subspaces@t1:1;\fi%
 \else \ifsingl@\subspaces@t3:5;\else\subspaces@t4:5;\fi \fi%
 \ifdoubl@ \multiply\baselineskip by 5%
 \divide\baselineskip by 4 \fi \unskip}
\def\subspaces@t#1:#2;{%
      \baselineskip = \normalbaselineskip%
      \multiply\baselineskip by #1 \divide\baselineskip by #2%
      \lineskip = \normallineskip%
      \multiply\lineskip by #1 \divide\lineskip by #2%
      \lineskiplimit = \normallineskiplimit%
      \multiply\lineskiplimit by #1 \divide\lineskiplimit by #2%
      \parskip = \normalparskip%
      \multiply\parskip by #1 \divide\parskip by #2%
      \abovedisplayskip = \normaldisplayskip%
      \multiply\abovedisplayskip by #1 \divide\abovedisplayskip by #2%
      \belowdisplayskip = \abovedisplayskip%
      \abovedisplayshortskip = \normaldispshortskip%
      \multiply\abovedisplayshortskip by #1%
	\divide\abovedisplayshortskip by #2%
      \belowdisplayshortskip = \abovedisplayshortskip%
      \advance\belowdisplayshortskip by \belowdisplayskip%
      \divide\belowdisplayshortskip by 2%
      \smallskipamount = \skipregister%
      \multiply\smallskipamount by #1 \divide\smallskipamount by #2%
      \medskipamount = \smallskipamount \multiply\medskipamount by 2%
      \bigskipamount = \smallskipamount \multiply\bigskipamount by 4 }
\def\normalbaselines{ \baselineskip=\normalbaselineskip%
   \lineskip=\normallineskip \lineskiplimit=\normallineskip%
   \iftwelv@\else \multiply\baselineskip by 4 \divide\baselineskip by 5%
     \multiply\lineskiplimit by 4 \divide\lineskiplimit by 5%
     \multiply\lineskip by 4 \divide\lineskip by 5 \fi }
\Twelvepoint  % That's the default
\interlinepenalty=50
\interfootnotelinepenalty=5000
\predisplaypenalty=9000
\postdisplaypenalty=500
\hfuzz=1pt
\vfuzz=0.2pt
%
%%%%%%%%%%%%%%%%%%%%%%%%%%%%%%%%%%%%%%%%%%%%%%%%%%%%%%%%%%%%%%%%%%%%%%%%
%
%   Next, I define output routines, footnotes & related stuff.
%
\def\pagecontents{%
   \ifvoid\topins\else\unvbox\topins\vskip\skip\topins\fi
   \dimen@ = \dp255 \unvbox255
   \ifvoid\footins\else\vskip\skip\footins\footrule\unvbox\footins\fi
   \ifr@ggedbottom \kern-\dimen@ \vfil \fi }
\def\makeheadline{\vbox to 0pt{ \skip@=\topskip
      \advance\skip@ by -12pt \advance\skip@ by -2\normalbaselineskip
      \vskip\skip@ \line{\vbox to 12pt{}\the\headline} \vss
      }\nointerlineskip}
\def\makefootline{\baselineskip = 1.5\normalbaselineskip
		 \line{\the\footline}}
\newif\iffrontpage
\newif\ifletterstyle
\newif\ifp@genum
\def\nopagenumbers{\p@genumfalse}
\def\pagenumbers{\p@genumtrue}
\pagenumbers
\newtoks\paperheadline
\newtoks\letterheadline
\newtoks\letterfrontheadline
\newtoks\lettermainheadline
\newtoks\paperfootline
\newtoks\letterfootline
\newtoks\date
\footline={\ifletterstyle\the\letterfootline\else\the\paperfootline\fi}
\paperfootline={\hss\iffrontpage\else\ifp@genum\tenrm\folio\hss\fi\fi}
\letterfootline={\hfil}
\headline={\ifletterstyle\the\letterheadline\else\the\paperheadline\fi}
\paperheadline={\hfil}
\letterheadline{\iffrontpage\the\letterfrontheadline
     \else\the\lettermainheadline\fi}
\lettermainheadline={\rm\ifp@genum page \ \folio\fi\hfil\the\date}
\def\monthname{\relax\ifcase\month 0/\or January\or February\or
   March\or April\or May\or June\or July\or August\or September\or
   October\or November\or December\else\number\month/\fi}
\date={\monthname\ \number\day, \number\year}
\countdef\pagenumber=1  \pagenumber=1
\def\advancepageno{\global\advance\pageno by 1
   \ifnum\pagenumber<0 \global\advance\pagenumber by -1
    \else\global\advance\pagenumber by 1 \fi \global\frontpagefalse }
\def\folio{\ifnum\pagenumber<0 \romannumeral-\pagenumber
	   \else \number\pagenumber \fi }
\def\footrule{\dimen@=\prevdepth\nointerlineskip
   \vbox to 0pt{\vskip -0.25\baselineskip \hrule width 0.35\hsize \vss}
   \prevdepth=\dimen@ }
\newtoks\foottokens
\foottokens={\Tenpoint\singlespace}
\newdimen\footindent
\footindent=24pt
\def\vfootnote#1{\insert\footins\bgroup  \the\foottokens
   \interlinepenalty=\interfootnotelinepenalty \floatingpenalty=20000
   \splittopskip=\ht\strutbox \boxmaxdepth=\dp\strutbox
   \leftskip=\footindent \rightskip=\z@skip
   \parindent=0.5\footindent \parfillskip=0pt plus 1fil
   \spaceskip=\z@skip \xspaceskip=\z@skip
   \Textindent{$ #1 $}\footstrut\futurelet\next\fo@t}
\def\Textindent#1{\noindent\llap{#1\enspace}\ignorespaces}
\def\footnote#1{\attach{#1}\vfootnote{#1}}

\let\footsymbol=\star
\newcount\lastf@@t	     \lastf@@t=-1
\newcount\footsymbolcount    \footsymbolcount=0
\newif\ifPhysRev
\def\footsymbolgen{\relax \ifPhysRev \iffrontpage \NPsymbolgen\else
      \PRsymbolgen\fi \else \NPsymbolgen\fi
   \global\lastf@@t=\pageno \footsymbol }
\def\NPsymbolgen{\ifnum\footsymbolcount<0 \global\footsymbolcount=0\fi
   {\iffrontpage \else \advance\lastf@@t by 1 \fi
    \ifnum\lastf@@t<\pageno \global\footsymbolcount=0
     \else \global\advance\footsymbolcount by 1 \fi }
   \ifcase\footsymbolcount \fd@f\star\or \fd@f\dagger\or \fd@f\ast\or
    \fd@f\ddagger\or \fd@f\natural\or \fd@f\diamond\or \fd@f\bullet\or
    \fd@f\nabla\else \fd@f\dagger\global\footsymbolcount=0 \fi }
\def\fd@f#1{\xdef\footsymbol{#1}}
\def\PRsymbolgen{\ifnum\footsymbolcount>0 \global\footsymbolcount=0\fi
      \global\advance\footsymbolcount by -1
      \xdef\footsymbol{\sharp\number-\footsymbolcount} }
\def\space@ver#1{\let\@sf=\empty \ifmmode #1\else \ifhmode
   \edef\@sf{\spacefactor=\the\spacefactor}\unskip${}#1$\relax\fi\fi}
\def\attach#1{\space@ver{\strut^{\mkern 2mu #1} }\@sf\ }
%
%%%%%%%%%%%%%%%%%%%%%%%%%%%%%%%%%%%%%%%%%%%%%%%%%%%%%%%%%%%%%%%%%%%%%%%%
%
%   Here come chapter, section, subsection & appendix macros.
%
\newcount\chapternumber	     \chapternumber=0
\newcount\sectionnumber	     \sectionnumber=0
\newcount\equanumber	     \equanumber=0
\let\chapterlabel=0
\newtoks\chapterstyle	     \chapterstyle={\Number}
\newskip\chapterskip	     \chapterskip=\bigskipamount
\newskip\sectionskip	     \sectionskip=\medskipamount
\newskip\headskip	     \headskip=8pt plus 3pt minus 3pt
\newdimen\chapterminspace    \chapterminspace=15pc
\newdimen\sectionminspace    \sectionminspace=10pc
\newdimen\referenceminspace  \referenceminspace=25pc
\def\chapterreset{\global\advance\chapternumber by 1
   \ifnum\equanumber<0 \else\global\equanumber=0\fi
   \sectionnumber=0 \makel@bel}
\def\makel@bel{\xdef\chapterlabel{%
\the\chapterstyle{\the\chapternumber}.}}
\def\sectionlabel{\number\sectionnumber \quad }
\def\alphabetic#1{\count255='140 \advance\count255 by #1\char\count255}
\def\Alphabetic#1{\count255='100 \advance\count255 by #1\char\count255}
\def\Roman#1{\uppercase\expandafter{\romannumeral #1}}
\def\roman#1{\romannumeral #1}
\def\Number#1{\number #1}
\def\unnumberedchapters{\let\makel@bel=\relax \let\chapterlabel=\relax
\let\sectionlabel=\relax \equanumber=-1 }
\def\titlestyle#1{\par\begingroup \interlinepenalty=9999
     \leftskip=0.02\hsize plus 0.23\hsize minus 0.02\hsize
     \rightskip=\leftskip \parfillskip=0pt
     \hyphenpenalty=9000 \exhyphenpenalty=9000
     \tolerance=9999 \pretolerance=9000
     \spaceskip=0.333em \xspaceskip=0.5em
     \iftwelv@\fourteenpoint\else\twelvepoint\fi
   \noindent #1\par\endgroup }
\def\spacecheck#1{\dimen@=\pagegoal\advance\dimen@ by -\pagetotal
   \ifdim\dimen@<#1 \ifdim\dimen@>0pt \vfil\break \fi\fi}
\def\chapter#1{\par \penalty-300 \vskip\chapterskip
   \spacecheck\chapterminspace
   \chapterreset \titlestyle{\chapterlabel \ #1}
   \nobreak\vskip\headskip \penalty 30000
   \wlog{\string\chapter\ \chapterlabel} }

\def\section#1{\par \ifnum\the\lastpenalty=30000\else
   \penalty-200\vskip\sectionskip \spacecheck\sectionminspace\fi
   \wlog{\string\section\ \chapterlabel \the\sectionnumber}
   \global\advance\sectionnumber by 1  \noindent
   {\caps\enspace\chapterlabel \sectionlabel #1}\par
   \nobreak\vskip\headskip \penalty 30000 }
\def\subsection#1{\par
   \ifnum\the\lastpenalty=30000\else \penalty-100\smallskip \fi
   \noindent\undertext{#1}\enspace \vadjust{\penalty5000}}

\def\undertext#1{\vtop{\hbox{#1}\kern 1pt \hrule}}
\def\APPENDIX#1#2{\par\penalty-300\vskip\chapterskip
   \spacecheck\chapterminspace \chapterreset \xdef\chapterlabel{#1}
   \titlestyle{APPENDIX #2} \nobreak\vskip\headskip \penalty 30000
   \wlog{\string\Appendix\ \chapterlabel} }
\def\Appendix#1{\APPENDIX{#1}{#1}}
\def\appendix{\APPENDIX{A}{}}
%
%%%%%%%%%%%%%%%%%%%%%%%%%%%%%%%%%%%%%%%%%%%%%%%%%%%%%%%%%%%%%%%%%%%%%%%%
%
%   Here come macros for equation numbering.
%
\def\eqname#1{\relax \ifnum\equanumber<0
     \xdef#1{{\rm(\number-\equanumber)}}\global\advance\equanumber by -1
    \else \global\advance\equanumber by 1
      \xdef#1{{\rm(\chapterlabel \number\equanumber)}} \fi}

\def\eqn#1{\eqno\eqname{#1}#1}

\def\eqinsert#1{\noalign{\dimen@=\prevdepth \nointerlineskip
   \setbox0=\hbox to\displaywidth{\hfil #1}
   \vbox to 0pt{\vss\hbox{$\!\box0\!$}\kern-0.5\baselineskip}
   \prevdepth=\dimen@}}
%

%

%

%
%%%%%%%%%%%%%%%%%%%%%%%%%%%%%%%%%%%%%%%%%%%%%%%%%%%%%%%%%%%%%%%%%%%%%%%%
%   Here come items and lists
%
\def\GENITEM#1;#2{\par \hangafter=0 \hangindent=#1
    \Textindent{$ #2 $}\ignorespaces}
\outer\def\newitem#1=#2;{\gdef#1{\GENITEM #2;}}
\newdimen\itemsize		  \itemsize=30pt
\newitem\item=1\itemsize;
\newitem\sitem=1.75\itemsize;	  
\newitem\ssitem=2.5\itemsize;	  
\outer\def\newlist#1=#2&#3&#4;{\toks0={#2}\toks1={#3}%
   \count255=\escapechar \escapechar=-1
   \alloc@0\list\countdef\insc@unt\listcount	 \listcount=0
   \edef#1{\par
      \countdef\listcount=\the\allocationnumber
      \advance\listcount by 1
      \hangafter=0 \hangindent=#4
      \Textindent{\the\toks0{\listcount}\the\toks1}}
   \expandafter\expandafter\expandafter
    \edef\c@t#1{begin}{\par
      \countdef\listcount=\the\allocationnumber \listcount=1
      \hangafter=0 \hangindent=#4
      \Textindent{\the\toks0{\listcount}\the\toks1}}
   \expandafter\expandafter\expandafter
    \edef\c@t#1{con}{\par \hangafter=0 \hangindent=#4 \noindent}
   \escapechar=\count255}
\def\c@t#1#2{\csname\string#1#2\endcsname}
\newlist\point=\Number&.&1.0\itemsize;
\newlist\subpoint=(\alphabetic&)&1.75\itemsize;
\newlist\subsubpoint=(\roman&)&2.5\itemsize;
%

%
%%%%%%%%%%%%%%%%%%%%%%%%%%%%%%%%%%%%%%%%%%%%%%%%%%%%%%%%%%%%%%%%%%%%%%%%
%
%   Here come macros for references, figures & tables.
%
\newcount\referencecount     \referencecount=0
\newif\ifreferenceopen	     \newwrite\referencewrite
\newtoks\rw@toks
\def\NPrefmark#1{\attach{\scriptscriptstyle [ #1 ] }}
\let\PRrefmark=\attach
\def\refmark#1{\relax\ifPhysRev\PRrefmark{#1}\else\NPrefmark{#1}\fi}
\def\refend{\refmark{\number\referencecount}}
\newcount\lastrefsbegincount \lastrefsbegincount=0
\def\refsend{\refmark{\count255=\referencecount
   \advance\count255 by-\lastrefsbegincount
   \ifcase\count255 \number\referencecount
   \or \number\lastrefsbegincount,\number\referencecount
   \else \number\lastrefsbegincount-\number\referencecount \fi}}
\def\refch@ck{\chardef\rw@write=\referencewrite
   \ifreferenceopen \else \referenceopentrue
   \immediate\openout\referencewrite=reference.aux \fi}
%
% In \obeyendofline, we say `\let^^M=\relax
{\catcode`\^^M=\active % these lines must end with %
  \gdef\obeyendofline{\catcode`\^^M\active \let^^M\ }}%
%
% In \ignoreendofline, we say `\let^^M=\relax
{\catcode`\^^M=\active % these lines must end with %
  \gdef\ignoreendofline{\catcode`\^^M=5}}
{\obeyendofline\gdef\rw@start#1{\def\t@st{#1} \ifx\t@st\blankend%
\endgroup \@sf \relax \else \ifx\t@st\bl@nkend \endgroup \@sf \relax%
\else \rw@begin#1
\backtotext
\fi \fi } }
{\obeyendofline\gdef\rw@begin#1
{\def\n@xt{#1}\rw@toks={#1}\relax%
\rw@next}}
\def\blankend{}
{\obeylines\gdef\bl@nkend{
}}
\newif\iffirstrefline  \firstreflinetrue
\def\rwr@teswitch{\ifx\n@xt\blankend \let\n@xt=\rw@begin %
 \else\iffirstrefline \global\firstreflinefalse%
\immediate\write\rw@write{\noexpand\obeyendofline \the\rw@toks}%
\let\n@xt=\rw@begin%
      \else\ifx\n@xt\rw@@d \def\n@xt{\immediate\write\rw@write{%
	\noexpand\ignoreendofline}\endgroup \@sf}%
	     \else \immediate\write\rw@write{\the\rw@toks}%
	     \let\n@xt=\rw@begin\fi\fi \fi}
\def\rw@next{\rwr@teswitch\n@xt}
\def\rw@@d{\backtotext} \let\rw@end=\relax
\let\backtotext=\relax

\newdimen\refindent	\refindent=30pt
\def\refitem#1{\par \hangafter=0 \hangindent=\refindent \Textindent{#1}}
\def\REFNUM#1{\space@ver{}\refch@ck \firstreflinetrue%
 \global\advance\referencecount by 1 \xdef#1{\the\referencecount}}
\def\refnum#1{\space@ver{}\refch@ck \firstreflinetrue%
 \global\advance\referencecount by 1 \xdef#1{\the\referencecount}\refend}

\def\REF#1{\REFNUM#1%
 \immediate\write\referencewrite{%
 \noexpand\refitem{#1.}}%
\begingroup\obeyendofline\rw@start}
\def\ref{\refnum\?%
 \immediate\write\referencewrite{\noexpand\refitem{\?.}}%
\begingroup\obeyendofline\rw@start}
\def\Ref#1{\refnum#1%
 \immediate\write\referencewrite{\noexpand\refitem{#1.}}%
\begingroup\obeyendofline\rw@start}
\def\REFS#1{\REFNUM#1\global\lastrefsbegincount=\referencecount
\immediate\write\referencewrite{\noexpand\refitem{#1.}}%
\begingroup\obeyendofline\rw@start}
\newcount\figurecount	  \figurecount=0
\newif\iffigureopen	  \newwrite\figurewrite
\def\figch@ck{\chardef\rw@write=\figurewrite \iffigureopen\else
   \immediate\openout\figurewrite=figures.aux
   \figureopentrue\fi}
\def\FIGNUM#1{\space@ver{}\figch@ck \firstreflinetrue%
 \global\advance\figurecount by 1 \xdef#1{\the\figurecount}}
\def\FIG#1{\FIGNUM#1
   \immediate\write\figurewrite{\noexpand\refitem{#1.}}%
   \begingroup\obeyendofline\rw@start}
\def\fig{\FIGNUM\? fig.~\?%
\immediate\write\figurewrite{\noexpand\refitem{\?.}}%
\begingroup\obeyendofline\rw@start}
\def\figure{\FIGNUM\? figure~\?
   \immediate\write\figurewrite{\noexpand\refitem{\?.}}%
   \begingroup\obeyendofline\rw@start}
\def\Fig{\FIGNUM\? Fig.~\?%
\immediate\write\figurewrite{\noexpand\refitem{\?.}}%
\begingroup\obeyendofline\rw@start}
\def\Figure{\FIGNUM\? Figure~\?%
\immediate\write\figurewrite{\noexpand\refitem{\?.}}%
\begingroup\obeyendofline\rw@start}
\newcount\tablecount	 \tablecount=0
\newif\iftableopen	 \newwrite\tablewrite
\def\tabch@ck{\chardef\rw@write=\tablewrite \iftableopen\else
   \immediate\openout\tablewrite=tables.aux
   \tableopentrue\fi}
\def\TABNUM#1{\space@ver{}\tabch@ck \firstreflinetrue%
 \global\advance\tablecount by 1 \xdef#1{\the\tablecount}}
\def\TABLE#1{\TABNUM#1
   \immediate\write\tablewrite{\noexpand\refitem{#1.}}%
   \begingroup\obeyendofline\rw@start}
\def\Table{\TABNUM\? Table~\?%
\immediate\write\tablewrite{\noexpand\refitem{\?.}}%
\begingroup\obeyendofline\rw@start}
%

%
%%%%%%%%%%%%%%%%%%%%%%%%%%%%%%%%%%%%%%%%%%%%%%%%%%%%%%%%%%%%%%%%%%%%%%%%
%
%   Here come macros for memos & letters.
%
\def\masterreset{\global\pagenumber=1 \global\chapternumber=0
   \global\equanumber=0 \global\sectionnumber=0
   \global\referencecount=0 \global\figurecount=0 \global\tablecount=0 }
\def\FRONTPAGE{\ifvoid255\else\vfill\penalty-2000\fi
      \masterreset\global\frontpagetrue
      \global\lastf@@t=0 \global\footsymbolcount=0}

\def\paperstyle{\letterstylefalse\normalspace\papersize}
\def\letterstyle{\letterstyletrue\singlespace\lettersize}
%
%\def\papersize{\hsize=35pc\vsize=50pc\hoffset=1pc\voffset=6pc
%		\skip\footins=\bigskipamount}
% 6/24/1986 two lines above was modified by J.CHIBA as below
\def\papersize{\hsize=35pc\vsize=50pc\hoffset=1cm\voffset=0pc
	       \skip\footins=\bigskipamount}
\def\lettersize{\hsize=6.5in\vsize=8.5in\hoffset=0.48in\voffset=1in
   \skip\footins=\smallskipamount \multiply\skip\footins by 3 }
\paperstyle   %  This is the default
%
% % % % % % % % % % % % % % % % % % % % % % % % % % % % % % % % % % % %
%
\def\MEMO{\letterstyle\FRONTPAGE \letterfrontheadline={\hfil}
    \line{\quad\fourteenrm MEMORANDUM\hfil\twelverm\the\date\quad}
    \medskip \memod@f}

\def\memit@m#1{\smallskip \hangafter=0 \hangindent=1in
      \Textindent{\caps #1}}
\def\memod@f{\xdef\to{\memit@m{To:}}\xdef\from{\memit@m{From:}}%
     \xdef\topic{\memit@m{Topic:}}\xdef\subject{\memit@m{Subject:}}%
     \xdef\rule{\bigskip\hrule height 1pt\bigskip}}
\memod@f
\newskip\lettertopfil
\lettertopfil = 0pt plus 1.5in minus 0pt
\newskip\letterbottomfil
\letterbottomfil = 0pt plus 2.3in minus 0pt
\newskip\spskip \setbox0\hbox{\ } \spskip=-1\wd0
\def\addressee#1{\medskip\rightline{\the\date\hskip 30pt}  \bigskip
 \vskip\lettertopfil
   \ialign to\hsize{\strut ##\hfil\tabskip 0pt plus \hsize \cr #1\crcr}
   \medskip\noindent\hskip\spskip}
\newskip\signatureskip	     \signatureskip=40pt
\def\signed#1{\par \penalty 9000 \bigskip \dt@pfalse
  \everycr={\noalign{\ifdt@p\vskip\signatureskip\global\dt@pfalse\fi}}
  \setbox0=\vbox{\singlespace \halign{\tabskip 0pt \strut ##\hfil\cr
   \noalign{\global\dt@ptrue}#1\crcr}}
  \line{\hskip 0.5\hsize minus 0.5\hsize \box0\hfil} \medskip }

\def\endletter{\ifnum\pagenumber=1 \vskip\letterbottomfil\supereject
\else \vfil\supereject \fi}
\newbox\letterb@x
\def\lettertext{\par\unvcopy\letterb@x\par}
\def\multiletter{\setbox\letterb@x=\vbox\bgroup
      \everypar{\vrule height 1\baselineskip depth 0pt width 0pt }
      \singlespace \topskip=\baselineskip }
\def\letterend{\par\egroup}
%
%%%%%%%%%%%%%%%%%%%%%%%%%%%%%%%%%%%%%%%%%%%%%%%%%%%%%%%%%%%%%%%%%%%%%%%
%
%   Here come macros for title pages.
%
\newskip\frontpageskip
\newtoks\pubtype
\newtoks\Pubnum
\newtoks\pubnum
\newif\ifp@bblock  \p@bblocktrue
\def\PH@SR@V{\doubl@true \baselineskip=24.1pt plus 0.2pt minus 0.1pt
	     \parskip= 3pt plus 2pt minus 1pt }
\def\PHYSREV{\paperstyle\PhysRevtrue\PH@SR@V}
\def\titlepage{\FRONTPAGE\paperstyle\ifPhysRev\PH@SR@V\fi
   \ifp@bblock\p@bblock\fi}
\def\nopubblock{\p@bblockfalse}
\def\endpage{\vfil\break}
\frontpageskip=1\medskipamount plus .5fil
\pubtype={ }
\newtoks\publevel
\publevel={Report}   % The alternatives are Internal and Preprint
\Pubnum={ }
%\pubnum={0000}
%
\def\p@bblock{\begingroup \tabskip=\hsize minus \hsize
   \baselineskip=1.5\ht\strutbox \topspace-2\baselineskip
   \halign to\hsize{\strut ##\hfil\tabskip=0pt\crcr
   \the\date\cr}\endgroup}
\def\title#1{\vskip\frontpageskip \titlestyle{#1} \vskip\headskip }
\def\author#1{\vskip\frontpageskip\titlestyle{\twelvecp #1}\nobreak}

\def\address#1{\par\kern 5pt\titlestyle{\twelvepoint\it #1}}
\def\andaddress{\par\kern 5pt \centerline{\sl and} \address}

% 6/24/1986 two lines below were added by H.Mawatari

%
\def\abstract{\vskip\frontpageskip\centerline{\fourteenrm ABSTRACT}
	      \vskip\headskip }

%
%
%%%%%%%%%%%%%%%%%%%%%%%%%%%%%%%%%%%%%%%%%%%%%%%%%%%%%%%%%%%%%%%%%%%%%%%%
%   Miscellaneous macros
%

\def\\{\relax\ifmmode\backslash\else$\backslash$\fi}
\def\globaleqnumbers{\relax\if\equanumber<0\else\global\equanumber=-1\fi}

\def\journal#1&#2(#3){\unskip, \sl #1~\bf #2 \rm (19#3) }

\def\topspace{\hrule height 0pt depth 0pt \vskip}

\def\ket#1{\left\vert #1\right\rangle}

\let\int=\intop		\let\oint=\ointop
\def\prop{\mathrel{{\mathchoice{\pr@p\scriptstyle}{\pr@p\scriptstyle}{
		\pr@p\scriptscriptstyle}{\pr@p\scriptscriptstyle} }}}
\def\pr@p#1{\setbox0=\hbox{$\cal #1 \char'103$}
   \hbox{$\cal #1 \char'117$\kern-.4\wd0\box0}}
\def\lsim{\mathrel{\mathpalette\@versim<}}
\def\gsim{\mathrel{\mathpalette\@versim>}}
\def\@versim#1#2{\lower0.2ex\vbox{\baselineskip\z@skip\lineskip\z@skip
  \lineskiplimit\z@\ialign{$\m@th#1\hfil##\hfil$\crcr#2\crcr\sim\crcr}}}
%
% % % % % % % % % % % % % % % % % % % % % % % % % % % % % % % % % % % %
%
%   Finally, some bug fixings.
%
\let\sec@nt=\sec
\def\sec{\relax\ifmmode\let\n@xt=\sec@nt\else\let\n@xt\section\fi\n@xt}
\def\obsolete#1{\message{Macro \string #1 is obsolete.}}
\def\firstsec#1{\obsolete\firstsec \section{#1}}
\def\firstsubsec#1{\obsolete\firstsubsec \subsection{#1}}
\def\thispage#1{\obsolete\thispage \global\pagenumber=#1\frontpagefalse}
\def\thischapter#1{\obsolete\thischapter \global\chapternumber=#1}
\def\nextequation#1{\obsolete\nextequation \global\equanumber=#1
   \ifnum\the\equanumber>0 \global\advance\equanumber by 1 \fi}
\def\BOXITEM{\afterassigment\B@XITEM\setbox0=}
\def\B@XITEM{\par\hangindent\wd0 \noindent\box0 }
%

%%%%%%%%%%%%%%%%%%%%%%%%%%%%%%%%%%%%%%%%%%%%%%%%%%%%%%%%%%%%%%%%%%%%%%%%
%   That's about it
%
\catcode`@=12 % at signs are no longer letters
\message{ by V.K.}
\def\bracket#1#2#3{\Big\langle{#1}\Big\vert{#2}\Big\vert{#3}\Big\rangle}

\def\nint{\int\nolimits}
\def\noint{\oint\nolimits}
%
% End of Macro
%
%
%
\title{ Equation of Motion for a Spin Vortex and
Geometric Force }
\author{Hiroshi Kuratsuji}
\address{Department of Physics, Ritsumeikan University-BKC,
               Kusatsu City 525-77, Japan}

\author{Hiroyuki Yabu}
\address{Department of Physics, Tokyo Metropolitan University,
               Hachioji, Tokyo 192, Japan}
\abstract{
The Hamiltonian equation of motion
is studied for a vortex occuring in 2-dimesional Heisenberg
ferromagnet of anisotropic type by starting with the
effective action for the spin field formulated by the
Bloch (or spin) coherent state. The resultant equation
shows the existence of a geometric force that
is analogous to the so-called Magnus force in superfluid.
This specific force plays a significant role for a quantum
dynamics for a single vortex, e.g, the determination  of the
bound state of the vortex trapped by a pinning force
arising from the interaction of the vortex with
an impurity. }

\endpage

%
%
%%%%%%%%%%
%
%%%%%%%%%%
%
%

\chapter{Introduction}

The quantum vortices in superfluids is one of most attractive
subjects in condensed matter physics.
Among many aspects in vortex phenomena,
the dynamics of many vortices has been developed by
starting with the relevant assumption on the boson superfluid
\Ref\Fetter{A.Fetter, Phys.Rev.{\bf 162}(1967)143;
J.Creswick and H.L.Morisson, \hfill\break
Phys.Lett.{\bf 76A}
(1980)267.}
,which reproduced the well known form for the Hamiltonian
equation for the assembly of vortices
\Ref\Lamb{
H.Lamb, Hydrodynamics, Cambridge, 1932;
L.Onsager,Nuovo Cimento \hfill\break
Suppl.6 (1949)279.}
.. The quantization based on the Hamiltonian equation
has also been studied.
\Ref\Chiao{
R.Y.Chiao, A. Hansen and A.A.Mouthrop,
Phys.Rev.Lett.{\bf 54}(1985)1339.}
Recently, a refined formulation  has been given for the
quatum treatment for superfluid vortex in the framework of the
generalized Hamiltonian dynmamics starting
with the the Landau-Ginzburg action
\Ref\Kura{
H.Kuratsuji, Phys. Rev. Lett. {\bf 68} (1992) 1746.}
.. Besides the superfluid vortex, the other types of quantum votex
has been interested for some time, typically, the vortex
in the Heisenberg ferromagnet~~(see for examle,
\Ref\Bishop{
M.E.Gouvea, G.M.Wysin, A.Bishop, and F.G.Mertens,Phys.Rev.
{\bf B39}\hfill\break
(1989)11840 and references therein. }
). The occurence of vortex in ferromagnet is quite natural, if one notes
a close resemblance between the superfluid He4 and the ferromagnet
as quantum condensate, especially, in the vicinity of the ground state.
Following the procedure developed in the superfluid,
the Hamiltonian dynamics has been studied for vortices occuring
in the 2-dimensional spin condensates.
\Ref\Ono{ H.Ono and H.Kuratsuji, Phys.Lett.{\bf A186}(1994)255.}

Apart from the quantum dynamics of assembly of vortices,
there has been long interest in the peculiar behavior of the
motion of a single vortex mainly relation to the type II superconductors.
\REF\Bardeen{ J. Bardeen and M.Stephen,
Phys. Rev. {\bf 140} (1965) 1197.}
\REF\Nozieres{ P. Nozieres and W.~F. Vinen,
Philos. Mag. {\bf 14} (1966) 667.}
\refmark{\Bardeen, \Nozieres}
One of the main object of this is concerning
the existence of a specific force called Magnus force.
This specific force is known to occur
when a vortex moves in the uniform stream and play
a role to explain some characteristic features of the type II
superconductors, (see e.g.
\Ref\Tilly{
D.R.Tilly and J.Tilly, Superfluidity and Superconductivity;
3-rd edition, \hfill\break
Adam Hilger, 1990.}
) The Magnus force is also known to play a crucial role
in the dissipative processes in the superfluid:
the smallness of the critical velocity,
the attenuation ratio of the second sound wave
in the rotating superfluid He${}^4$ and so on.
\Ref\Feynman{ R. Feynman,
Statistical Mechanics, W.A.Benjamin,  1972.}

The purpose of this paper is to put forward the equation of
motion for a spin vortex for the ferromagnetic system
within the Hamiltonian formulation of quantum vortex
previously developed.
\refmark{\Ono}
As a consequence of this, we naturally arrive at a force of
Magnus type. Indeed, if one considers the resemblance between the
superfluid He and the ferromagnet as quantum condensates,
it is natural to expect a realization of such an analogous force.
This force should be called the "geometric force",
which differs from the ordinary force derived from a potential
function. We also show the other type of force called the pinning
force, which comes from the interaction between a vortex and an
impurity immersed in condensate. It is shown that the effect of the
pinning force is realized by the bound state of the vortex trapped in
the pinning potential.

\chapter{Spin field Lagrangian}

Our starting point is the spin coherent state (orBloch state):
\Ref\KraIII{
in "Path Integrals and Coherent States of SU(2) and SU(1,1)",
A.Inomata, \hfill\break
H.Kuratsuji and C.Gerry, World Scientific, Singapore, 1992.}
The quantum state for the spin system can be described by an
infinite product of the Bloch state defined on each space point
(designated by discrete vector $ \vec n $ which is assumed to
take over to continous)
$$
\ket{\{z(x)\}}= \prod_{\vec n}\ket{z_{\vec n}}
\eqn\one
$$
where $ \vec n $ means the vector assigning the lattice point.
Each component is given by the SU(2) coherent state:
$$
\ket{z}={1 \over (1 + |z|^2)^J} \exp[z\hat J_{+}]\ket{0}
\eqn\two
$$
where $\ket{0} = \ket{J, -J}$ is the lowest state satisfying
${\hat J}_- \ket{0} =0$ and ${\hat J}_\pm$ are ladder operator
and $z$ takes any complex values. Using \one, we have
the action function
$$   S = \int \bracket{\{z(x)\}}
     {i\hbar{\partial \over \partial t} - \hat H}{\{z(x)\}}dt
       = \int {\cal L}d^2xdt
\eqn\three
$$
the Lagrangian density is given as
$$
{\cal L} = {iJ \hbar \over 2}{z^* {\dot z} -
            {\dot z}^* z \over 1 + |z|^2} - H(\{z(x)\},\{z(x)\}^{*})
\eqn\four
$$
Here note that the first term is a counterpart of the so-called
"geometric phase",
which is represented in terms of the overcomplete set:
\Ref\KraII{
H.Kuratsuji, Prog.Theor.Phys.{\bf 65}(1981)224;
Phys.Rev.Lett.{\bf 61}(1988)1687.}
$$
\Gamma= \oint \bracket{Z}{i\hbar{\partial \over \partial t}}{Z}dt
\eqn\five
$$
The quantization may be realized by the constructing the
propagator
$$
K = \bracket{{z(x)}}{\exp[-{iH\over \hbar}T]}{{z(x)}}
       = \int \exp[{i\over \hbar}\int {\cal L}dxdt]\prod_{x,t}d\mu[z(x,t)]
\eqn\fivex
$$

We shall now adopt the Heisenberg spin model for analyzing the
vortex dynamics.
As the Hamiltonian, we take the continuous version
of the nearest neigbour interaction of anisotropic type:
$$
{\hat H} = {g \over 2}\{(\nabla {\hat J}_1)^2
 + (\nabla {\hat J}_2)^2
 + \lambda(\nabla {\hat J}_3)^2\}
\eqn\six
$$
where the parameter means the degree of anisotropy,
which is assumed to be $ 0 < \lambda < 1 $, and this makes
the system to favour the planer spin configuration.
The expectation value becomes
$$
H = {g \over 2}\{(\nabla J_1)^2
 + (\nabla J_2)^2
 + \lambda(\nabla J_3)^2\}
\eqn\seven
$$
We have the more transparent form if we use
the angle variables via the stereographic projection,
$$
z = \tan {\theta \over 2} {\rm e}^{-i\phi}
\eqn\eight
$$
with $ (0 \leq \theta \leq \pi \, , ~0 \leq \phi \leq 2 \pi) $ ,
or the angular expression for the spin variables,
then it is also written as
$$
H = {g \over 2}J^2
\left\{(\cos^2\theta+ \lambda\sin^2\theta)
(\nabla \theta)^2 + \sin^2 \theta (\nabla \phi)^2 \right\}
\eqn\nine
$$
The second term may be called the "fluid kinetic energy":
$$
   T \equiv {1\over 2}gJ^2 \int \sin^2\theta (\nabla \phi)^2d^2x
\eqn\ten
$$
The variation principle results in the field equation for
the angle variable, namely,
$$
\delta \int [J\hbar(1 - \cos\theta)\dot\phi - H(\theta, \phi)]d^2xdt
=0
\eqn\eleven
$$
which yields
$$
\eqalign{
J\hbar\sin\theta{\partial \theta \over \partial t} & =
            -{\partial H \over \partial \phi}, \cr
J\hbar\sin\theta{\partial \phi \over \partial t} & =
            {\partial H \over \partial \theta } }
\eqn\twelve
$$

\chapter{Equation of Motion for a Spin Vortex}

We consider a static solution exhibiting
the vortex, which is given from a static form of the
variation equation, namely, the vortex solution is characterized by
choosing the azimuthial angle as
$$
\phi = \tan^{-1}{y \over x}
\eqn\thirt
$$
and futhermore the lateral angle $ \theta $ can be chosen as a
function of the radial variable $ r $ only. Hence,
$$
    H = \int \big[r(\cos^2\theta+ \lambda\sin^2\theta)
      ({d\theta \over dr})^2
                 + {1\over r}\sin^2\theta\big]dr
\eqn\fourt
$$
The equation for the profile function $ \theta $ is
derived from the condition such that $ H $ takes an extremum.
This may be solved by imposing the proper boundary condition
for $ \theta(r) $ such that the z-component spin $ J_3(x) $ directs
upward inside the core (of radius $ a $) and vanishes outside the core.
Physically speaking, this boundary condition may be considered
to be an idealization of the feature that the spin configuration
is planer at infinity.  Instead solving exactly, it may be possible to
similate the profile for $\theta(r) $; for example we can choose
$$
\theta(r)  = \cases{
    cr  & ($ 0 \leq r \leq a  \equiv {\pi \over 2c}$) \cr
    {\pi \over 2}    & ($ a \leq r$) \cr
}
\eqn\twthree
$$

We shall now turn to the dynamics for a single vortex.
We first adopt the polar form for the complex field,
namely, the stereographic projection given above.
The physical meaning of the angle $ \phi $ is crucial,
namely,  that is assumed to coincide with the polar angle in
the coordinate plane $(x,y)$, which is measured from the
vortex center coordinate $ {\bf X}(t) = (X(t), Y(t)) $:
$$
    \phi = \mu \tan^{-1}{y-Y(t) \over x-X(t)}
\eqn\fift
$$
Here the coefficient $ \mu $
stands for the vortex strength. Now, the canonical
term is expressed as
$$
{\mit L}_C = \int{J\hbar \over 2} (1 - \cos \theta)
{\dot \phi}d^2x
\eqn\sixt
$$
By making use of the chain rule,
$$
 {\partial \phi \over \partial t}
= {\partial \phi \over \partial {\bf X}}{d{\bf X} \over dt}
\eqn\sevent
$$
together with the fact that the gradient of the phase function gives the
velocity field, namely, if we note the phase is given as a function
of $ {\bf x} - {\bf X} $, we get
$$
    {\partial \phi \over \partial {\bf X}}
= \nabla \phi = {\bf v},
\eqn\eighteen
$$
where $ {\bf v}  $ denotes the velocity field coming
from the vortex we are concerned. Hence we get
for the canonical term
$$
 L_C= \int {J\hbar\over 2}(1 - \cos\theta)\nabla \phi
   \dot{\bf X}dx
\eqn\nint
$$
or which can be written as the differential one form:
$$
  \omega =  \int \eta {\bf v}\cdot d{\bf X}d^2x.
\eqn\twenty
$$
where the following quantity is defined ;
$$
\eta = {J\hbar  \over 2}(1 - \cos \theta )
\eqn\twone
$$
By using this notation, the boundary condition for $ \theta $
assigned in the above takes over to the profile function $ \eta $
in the canonical term: $ \eta \rightarrow 0 ({\rm for}~~r=0 ) $
and $ \eta \rightarrow {J\hbar \over 2}
({\rm for}~~r \rightarrow \infty) $. This feature
is similar to the vortex for bose fluid.
\refmark{\Kura}
Further we note the fact that the velocity field is written as
$$
    {\bf  v} = \mu {\bf k} \times \nabla\log\vert {\bf x}- {\bf X} \vert
\eqn\twtwo
$$

{\bf Canonical Term: }

We shall derive the Lagrangian for the motion of a vortex
which is described by the vortex center coordinate.
It is firstly noted that the contribution from the Hamiltonian,
\fourt, becomes constant and not relevant to the
dynamics of the vortex. So the term which we need is the
canonical term $ L_C$. Using the expressin \nint, we get
\Ref\Yabu{ H. Yabu and H. Kuratsuji, to be submitted.}
$$
\eqalign{
     L_C & = \int{d^2x} \eta({\bf x}-{\bf X})
                ({\bf v}({\bf x}
                     -{\bf X}) \cdot \dot{{\bf X}}) \cr
          & = I_1 \dot{X} +m I_2 \dot{Y},}
\eqn\twfour
$$
where $I_{\alpha} $ in \twfour~ is given by
$$
   I_{\alpha} = \int \eta {\bf v}_{\alpha} d^2x
\eqn\twfive
$$
The integrand of $I_\alpha$ does not decrease enough fast
at the large distance and the result depends on how to take the limit,
$r \rightarrow \infty$, in the integral boundary.
This may be handled by a sort of "regularization"
and some care should be taken to get the correct results.
Having carried out the regularization to evaluate the integral,
(see the appendix ), the final result for the canonical term
follows
$$
L_C ={ \eta_0\mu \over 2}(Y\dot X - X\dot Y)
\eqn\twsix
$$
If we note the boundary condition for $ \theta = {\pi \over 2} $
at $ r \rightarrow \infty $, the value of $ \eta $
is given by
$$  \eta_0 = {J\hbar \over 2}
\eqn\twsixa
$$
The formula \twsix~ is the main result of the present paper.

{\bf Pinning Effect by Impurity:}

    Next we consider the role of the impurity effect
for a vortex within the effective theoretical approach. It
may be plausible that the interaction comes from the magnetic origin.
The simplest choice satisfying this criterion
may be given as follows:
$$
L_{pin} = G \int{d^2x}J_3(x)s_3(x),
\eqn\twseven
$$
where $G$ is a coupling constant. Here $ s_3(x) $ represents the
spin that is carried by a magnetic impurity.
     Let us consider the case that one magnetic impurity is
located at ${\bf Y}$, and has the well-localized density distribution.
Then we can take the delta function approximation for
such a spin disctribution
$$
     s_3(x) =s_3\delta^2({\bf x}-{\bf Y}).
\eqn\tweight
$$
where $ s_3 = \pm {1\over 2} $. With the aid of \twthree,
the integral in \twseven~ is calculated easily
to result in the Lagrangian for the effective interaction
between the vortex and impurity:
$$
L_{impurity} ( \equiv ) U_{pin} = \cases{
     G s_3\cos[{\pi \over 2a}| {\bf X} -{\bf Y}|]
                       & ($ |{\bf X} -{\bf Y}|<a $) \cr
           0 & ($|{\bf X} -{\bf Y}|>a $) \cr }
\eqn\twnine
$$
Here, if the coupling constant $ G $ is positive, the sign of
spin $ s_3 $ is chosen such that the potential of the first
half of \twnine~becomes attractive.

{\bf Hamiltonian Equation of Motion:}

%
%%%
%
    Now combning \twsix,  \twnine, the effective lagrangian
for one vortex in the presence of impurity becomes
$$
     L_{eff} ={ \eta_0\mu \over 2}
      (Y\dot X - X\dot Y) - H_{eff}
\eqn\thirty
$$
Here the second term is the Hamiltonian
for the single vortex, which is nothing but the
pinning potential;  $ H_{eff} = U_{pin} $.
The equation of motion for it is obtained as the Euler-Lagrange equation:
$$
     \dot{X} =   {\partial H_{eff} \over \partial Y},
     \dot{Y} = - {\partial H_{eff} \over \partial X},
\eqn\thirtyone
$$
which leads to
$$ {\mu \eta_0 \over 2} ({\bf k} \times \dot{{\bf X}})
     =   {\partial U_{pin} \over \partial {\bf X}}.
\eqn\thirtytwo
$$
where the vector $ {\bf k} $ denotes the unit vector
perpendicular to the (x,y) plane.\hfill\break
\thirtyone~ can be regarded as a special case of the canonical equation of
motion,
\refmark{\Kura, \Ono}
where the pair$ (X,Y) $ should be regarded as a canonical
pair each other.

 From the physical point of view, the first term is nothing but the
"geometric force", while the second term represents the pinning force
derived from the pinning potential, which is the
usual force derived from the potential function.

{\bf Bound State by Pinning Potential}

The motion of vortex can be quantized if we note the
vortex coordinate $ (X,Y) $ is a canonical pair,
which leads to the bound state spectra formed of the vortex
trapped by the pinning potential. We estimate it
by using the Bohr-Sommerfeld (BS) quantization.
The BS rule is given by
$$   {\mu\eta_0 \over 2}\noint_C (XdY - YdX) = 2\pi n\hbar
\eqn\thirtyfour
$$
Here C means a loop which is defined as $ U_{pin} = E $.
For simplicity, the center of impurity is assumed to be
placed at the origin $ {\bf Y } = 0 $, and $ Gs_3 < 0 $,
so the energy contour is given by
$$    Gs_3\cos{\pi \over 2a}\vert {\bf X} \vert = E
\eqn\thirtyfive
$$
 From this form, the loop C becomes a circle;
$ \vert {\bf X} \vert = \sqrt{X^2 + Y^2} \equiv \rho $,
hence the quantization rule turs out to be
$$   {\mu\eta_0 \over 2}\pi\rho^2 = 2\pi n\hbar
\eqn\thirtysix
$$
with $ n = {\rm integer} $. Thus we get the energy spectrum
$$   E_n = Gs_3\cos\big[{\pi \over a}\sqrt{{n\hbar \over \mu\eta_0}}\big]
\eqn\thirtyseven
$$
The critical bound state is limited by $ E_n = 0 $,
which means the inequality;
$$  {\pi \over a}\sqrt{{n\hbar \over \mu\eta_0}} \leq {\pi \over 2}
$$
This leads to the condition for the quantum number $ n $:
$$  n\hbar \leq {\mu\eta_0a^2 \over 4}
\eqn\thityeight
$$

\chapter{Summary}

We have studied a possible occurence of the geometric force
in a magnetic condensate. This force is analogous to the Magnus
force in ordinary superfluids.
The characteristic property is the nature of the
"transversality", so to speak, since the force is
perpendicular to the velocity of "particle(vortex)", which suggests that
the force does not attribute to the energy dissipation.
This feature is a characteristics of
the Lorentz force, so the geometric force is a sort of the
Lorentz force. However, it should be noted that the analogy with the
Magnus force is not complete, since in the magetic condensate
we have no supercurrent as in the case of the superfluids.

 From the above derivation, the geometric force is attributed to
the canonical term, which arises from the geometric phase.
 From the formulation point of view,
the geometric force may be regarded as a special
case of the pervious treatment of the many vortex dynamics.
However, the effective Lagrangian for the single vortex
can naturally incorporate the effect of pinning force,
if we include the interaction with the magnetic
impurities immersed in the magnetic substance.
Indeed, we have shown that by using the Bohr-Sommerfeld quantization
the geometric force results in the
bound state of a vortex which is captured by a pinning potential.
Apart from such a potential problem, the geomtric force
would play a role for an estimate of an effect of dynamical perturbation
acting for the vortex motion. The details of this will be given
eleswhere.

\appendix

     In this appendix, we evaluate the integral
which appears in the canonical term,
$$
     I_\alpha =\int \eta v_{\alpha}d^2x .
\eqn\a
$$

     We define two discs as
$$
\eqalign{
     D_A & =\{ {\bf x} \;;\; |{\bf x}| \leq R_A \},   \cr
     D_a & =\{ {\bf x} \;;\; |{\bf x}-{\bf X}| \leq a \},}
\eqn\b
$$
where $a$ is the size of vortex defined in and $R_A$ should be
taken large enough for $D_A$ to include $D_A$.
Let's define $V(a,A)$ as the region,
$$
     V(a,A) =D_A -D_a.
\eqn\c
$$
     The integral $I_\alpha$ is indefinite at
$r \rightarrow \infty$, and we take the regularization
to get the finite result;  the integral region is restricted
to the finite region $V(a,A)$ (cut-off),
and the infinite limit is taken
$$
     I_\alpha =I_\alpha(a) +\lim_{R \rightarrow \infty} I_\alpha(R),
\eqn\d
$$
where each integral is written as
$$
\eqalign{
    I_\alpha(a) & = \int_{D_a} d^2x\eta v_{\alpha},  \cr
        I_\alpha(R) & = \int_{V(a;R)} d^2x \eta v_{\alpha} }.
\eqn\e
$$

    By using the polar coordinates around $x =X$,
the integral $I_\alpha(a)$ can be evaluated:
$$
\eqalign{
     I_\alpha(a) &= {\int_{D_a} r dr d\theta}
                       {\eta_0 \over a} r {\mu \over 2\pi}
                       {\epsilon_{\alpha\beta} x_\beta \over r^2},\cr
                 & = {\eta_0 \mu \over 2\pi a} \epsilon_{\alpha\beta}
                     {\int_0^a r dr} {\int_0^{2\pi} \hat{x}_\beta d\theta}
                    ={\eta_0 \mu \over 2\pi a} \epsilon_{\alpha\beta}
                     {a^2 \over 2} \cdot 0 =0, }
\eqn\f
$$
where $\hat{x}_\alpha =x_\alpha /r$. With the aid of the Stokes theorem,
the integral $I_\alpha(R)$ can be represented with the line integral:
$$
\eqalign{
     I_\alpha(R) &=-\epsilon_{\alpha\beta} \eta_0 \int_{V(a;R)}{d^2x}
                                            {\partial_\beta} G_V(x;X),\cr
                 &=-\epsilon_{\alpha\beta} \eta_0
                    \int_{{\partial D_A} +{\partial D_a}}
                                            ds_\beta G_V(x;X), }
\eqn\g
$$
Furthermore, by using the polar coordinate around $x=X$, the integral
on $\partial{D_a}$ becomes
$$
     \int_{\partial D_a}{ds_\beta} G_V(x;X)
          =-{\mu \over 2\pi} \log{a} {\int ds_\beta} =0,
\eqn\h
$$
and, for around $x=0$, the integral on $\partial{D_R}$ is evaluated
to be
$$
\eqalign{
     \int_{\partial D_R}{ds_\beta} G_V(x;X)
          &= -{\mu \over 2\pi} \int{ds_\beta}
                \left[ \log{R} +{({\bf x} {\bf X}) \over R^2}
                               +{\cal O}(R^{-2}) \right] \cr
          &= -{\mu \over \pi} {1 \over R^2} \int{ds_\beta} ({\bf x} {\bf X})
            +{\cal O}(R^{-2})  \cr
          &= {\mu \over 2\pi} X_\alpha \int{d\theta} \hat{x}_\alpha
             \hat{x}_\beta+{\cal O}(R^{-2})
           ={\mu \over 2} X_\beta +{\cal O}(R^{-2}). }
\eqn\i
$$

     With combining \h  and \i, we get the final result
for $I_\alpha^V$:
$$
I_\alpha ={\eta_0 \mu \over 2\pi} \epsilon_{\alpha\beta} X_\beta.
\eqn\j
$$
%
%%%%%%%%%%
%

\endpage
\par \penalty-400 \vskip\chapterskip
%  \spacecheck\referenceminspace \closeout\referencewrite
% 9/24/1986 The above 1 line was changed as below by H.Mawatari
   \spacecheck\referenceminspace \immediate\closeout\referencewrite
   \referenceopenfalse
   \line{\fourteenrm\hfil REFERENCES\hfil}\vskip\headskip
   \input reference.aux
   
\bye